# Evaluating Open Access Advantages for Citations and Altmetrics (2011-21): A Dynamic and Evolving Relationship

**Mike Taylor, BSc (Hons)**

University of Wolverhampton; Digital Science

## Abstract

Differences between the impacts of Open Access (OA) and non-OA research have been observed over a wide range of citation and altmetric indicators, usually finding an Open Access Advantage (OAA) within specific fields. However, science-wide analyses covering multiple years, indicators and disciplines are lacking. Using citation counts and six altmetrics for 38.7M articles published 2011-21, we compare OA and non-OA papers. The results show that there is no universal OAA across all disciplines or impact indicators: the OAA for citations tends to be lower for more recent papers, whereas the OAAs for news, blogs and Twitter are consistent across years and unrelated to volume of OA publications, whereas the OAAs for Wikipedia, patents and policy citations are more complex. These results support different hypotheses for different subjects and indicators. The evidence is consistent with OA accelerating research impact in the Medical & Health Sciences, Life Sciences and the Humanities; that increased visibility or discoverability is a factor in promoting the translation of research into socio-economic impact; and that OA is a factor in growing online engagement with research in some disciplines. OAAs are therefore complex, dynamic, multi-factorial and require considerable analysis to understand.

## Introduction

The last quarter century has seen revolutions in research communication driven by the adoption of the Internet as a means of diffusing and discussing scholarly findings. Amongst these changes, the rise of freely available – or Open Access (OA) – publications, and the phenomena of altmetrics – are two of the most significant. Altmetrics, or alternative metrics, are counts of mentions of academic research on various websites and Internet services. They are used as an adjunct to academic citations to understand the dissemination and impact of research through social networks, grey literature, and other non-scholarly venues, such as news and blogs.

Various studies have shown that OA publications tend to be more cited and mentioned than their non-OA equivalents, a phenomenon known as the Open Access Advantage (OAA) (Wang et al., 2015a;Piwowar et al., 2018; Holmberg et al., 2019; Taylor, 2020). However, as discussed below, the relationship between these measurements and the overall rates of OA publication has not hitherto been reported at scale or over prolonged periods. Similarly, there has been insufficient research into the underlying causes of OAAs, and of their development over time.

It is vital that researchers, research managers and funders understand the implications of OA publishing for the impact and dissemination of their scholarly outputs as OA publishing can be expensive; consequently it is important to weigh the cost against the benefits and to understand some of the implications behind changes in publication policy.

## The Growth of Open Access, its Drivers and Goals

The goals of the OA and the related Open Science (OS) movements are to make scientific research accessible to all levels of society, including academic scholars, non-academic

2 | Evaluating Open Access Advantages (2011-21)

researchers and other members of the public (Suber, 2012). Additionally, OA and OS are intended to promote transparency, collaboration, reproducibility, and public engagement in science (UNESCO, 2022). OA publishing may also increase the speed of research (Woelfle et al., 2011). The transition towards OA has been challenging (Widmark & Hamrin, 2019), offering new problems to both academic institutions (Barbers et al., 2018) and publishers, who have had to adopt new business models (Inchcoombe et al., 2022). The OA movement began to gain adoption in 2001 (Suber, 2012), following on earlier preprint repositories, including RePEC and Arxiv (Ginsparg, 2021). The term "Open Access" encompasses several ways in which research publications can be open. These are usually identified by colours: Gold, Bronze, Green, Platinum / Diamond and Black.

- **Platinum / Diamond OA:** forms of Gold OA where no fee is required (Pearce, 2022)
- **Gold OA**: authors pay a fee to make content freely available on the journal's website, with an OA licence.
- **Bronze OA:** a fee may or may not be paid, however the publication is freely available on the journal's website without any appropriate OA licence (Piwowar et al., 2018).
- **Green OA:** applies when an article is published in a repository, whether as a pre-print or after being published in a journal – which itself may or may not be OA.
- **Black OA**: the free sharing of research contrary to its licence. This category is excluded from this study.

The study of OA adoption has been facilitated by the development of databases that record the OA status of research publications. The Directory of Open Access Journals (DOAJ) has been registering OA research for over two decades (*DOAJ at 20 – DOAJ*, n.d.) and Crossref permits publishers to publish relevant license information alongside publication metadata (*License Information - Crossref*, n.d.). Unpaywall, a service launched in 2017, brings together data from these - and additional - sources to facilitate discovery and analysis at scale (Else, 2018). Although the initial accuracy of Unpaywall varied by discipline (Sergiadis, 2019), by 2021, it was found to be adequate for research and its coverage is continuing to grow (Jahn et al., 2021).

Although the adoption of OA has been uneven across disciplines (Frantsvåg & Strømme, 2019; Laakso & Björk, 2022), journals (Ko et al., 2009) and countries (Simard et al., 2022), overall growth has been fast. In 2009, overall rates of 20% were reported (Björk et al., 2010), 45% in 2015 (Piwowar et al., 2018), and 54% in 2022 (Basson et al., 2022). This growth has been driven by policies and mandates that have emerged since the beginning of the millennium. In particular, Canada, the United Kingdom, the European Union and the USA have published policies aimed at driving OA adoption (Canadian Institutes of Health Research, 2006; European Commission 2008; National Institutes of Health, 2009; UK Government, 2012;). Many policies have stipulated an embargo, whereby an article can be closed for a certain period – typically 12 months – but thereafter be made OA – however, President Biden's White House Office of Science and Technology Policy (OSTP) has issued a new policy, whereby all US Government funded research must be OA throughout its lifecycle (Nelson, 2022). Another initiative, Plan S (Schiltz, 2018), has been organized by a group of governments, funders and other institutions (Farley & Aspaas, 2023), with the stated ambition of making all research publications funded or supported by them OA by 2021 (Coalition S, 2008). Growth in OA publishing in medical fields was dramatically stimulated by the widespread adoption of OA for COVID-19 throughout 2020-21 (Park et al., 2021).

There have been several criticisms of OA publishing, with researchers citing the potential growth in low quality journals (Beall, 2015) and the growing presence of predatory journals. The latter masquerade as legitimate journals, but exist solely to profit through author payments without providing the core quality control or archiving guarantees expected from academic journals (Knox, 2013). However, other research has suggested that the 'moral panic' about predatory publishers may be misplaced as citation rates tend to be low. Furthermore attitudes towards them may have racist overtones, and come from a position of established privilege (Houghton, 2022).

The cost of OA is enormous, with the annual cost of Article Processing Charges (APC) being estimated at $2B (Zhang et al., 2022), although OA agreements with commercial publishers may save money during the transition to OA (JISC, 2022). Thus, research managers need evidence of the value of OA to help them balance its costs and benefits.

## Metrics, Altmetrics and Understanding Research Impact

The study of research impact has its roots in the availability of computers for document processing (Garfield, 1954) with references being used to prepare citation indices (Garfield, 1955) for many decades. Current citation indices include the Web of Science (Birkle et al., 2020), Scopus, Google Scholar (Kulkarni et al., 2009) and Dimensions (Hook et al., 2018). The availability of these indices has led to the development of many metrics designed to reflect the productivity and scientific impact of research (Aksnes et al., 2019), such as the Field Weighted Citation Index (Moed et al., 2012). There have been many criticisms of these metrics, including for those that do not take into account variations between disciplines (Hicks et al., 2015).

From the mid-90s, the field of 'webometrics' (Almind & Ingwersen, 1997) emerged from the fields of bibliometrics and scientometrics. Pioneers in this field analysed hyperlinks, usage data and related content to understand the emerging online world of scholarly communication (Bar-Ilan, 2000; Thelwall, 2000). The term 'altmetrics' was introduced in 2010 to encompass a range of online venues where interaction with research publications was evident, including social networks, blogging platforms, news and Wikipedia (Priem et al., 2010). The goal was to develop indicators of non-academic impacts. The initially distinctive feature of altmetrics was its orientation towards social web services with Application Programming Interfaces (API), although the term 'altmetrics' now also encompasses traditional webometrics and quasi-academic attention from grey literature, such as policy and patent citations (Haunschild & Bornmann, 2017).

The multiple and varied data sources comprising altmetrics make them heterogeneous (Haustein, 2016). The extent and nature of the impact, if any, reflected by altmetrics varies by data sources, by discipline (Fang et al., 2020), over time (Taylor, 2023), and by region (Basson et al., 2022). Citation data shows similar features for scholarly impact (Waltman, 2016).

Both academic and non-academic audiences use social media to engage with research (Sugimoto & Larivière, 2017) and approximately half of the tweets linking to journal articles are from academics (Mohammadi et al., 2018). Research funders often require researchers to write an impact plan that covers non-academic impact (Bornmann, 2013), and to have impact management plans (Britt Holbrook & Frodeman, 2011). Studies of altmetrics have therefore assessed their value as evidence for social impact (Bornmann, 2014) and to analyse non-scholarly usage (Mohammadi et al., 2015). Citation-based indicators have been used in the assessment of research groups, departments, and institutions (Moed, 2017). The relationships between citations and altmetrics have been studied to understand their ability to predict future citation rates (Eysenbach, 2011; Thelwall, 2018) despite the not insignificant challenges that their heterogeneity involves (Thelwall, 2016).

## The Open Access Citation and Altmetrics Advantages

The hypothesised OAA may be divided into two related phenomena: the OA Citation Advantage (OACA) and the OA Altmetrics Advantage (OAAA). The OACA was reported to be approximately 25% shortly after the advent of OA publishing (i.e., OA articles receiving 25% more citations than non-OA), with Gold OA typically being cited at a higher rate than Green OA (Eysenbach, 2006). This estimate has since been reduced to an 18% advantage, largely driven by Green and Hybrid OA publications (Piwowar et al., 2018). A meta-analysis of 134 studies revealed a mixed picture, however, with 64 finding a general OACA, 32 finding an OACA within subsets of their data and 37 failing to find it (Langham-Putrow et al., 2021). A study of the citation behaviour of articles that switched OA status observed the effect to be more pronounced for articles that were already highly cited, when compared to equivalent non-OA articles (Ottaviani, 2016).

Altmetrics can also be used to compare OA and non-OA research outputs (Wilsdon et al., 2015). The presence of OAAA for the volume of attention on both Twitter and Mendeley for articles in a single hybrid journal has been reported (Adie 2014), along with an absence of an OAAA for blogs and news sources. The presence of an OAAA mean of 47% for Wikipedia citations has been reported at a journal-level (Teplitskiy et al 2017). The OAAA of five separate data sources has been observed to vary over time in a longitudinal study (Taylor, 2023).

A study of Finnish papers confirmed the existence of an OAAA for some fields and attention sources, but reported a *disadvantage* for other fields and attention sources (Holmberg, Hedman, Bowman, Didegah & Laakso 2020). Although this research analysed the most populous altmetric indicators in detail (Twitter and Mendeley), as well as citations from the Web of Science, it compounded other altmetric indicators (news, blogs, Wikipedia and Facebook). However, this research used the OA journal status, as defined by the DOAJ, meaning that OA articles in hybrid journals, and Green OA articles would have been treated as non-OA.

A widescale analysis of publications found that Green OA publications received more citations than other OA types, and non-OA papers, but that Gold OA generally outperformed other OA types across a range of four altmetrics indicators (Twitter, News, Facebook and blogs), and that Green OA slightly outperformed other OA types for patent citations (Nishioka & Farber, 2020); however, this paper is limited to research publications with altmetrics data, and thus excludes the majority of publications.

There are several studies into narrowly defined topics. OA research in human electrophysiology gets up to 21% more citations and 39% more Altmetrics mentions that non-OA (Clayson et al, 2021); OA outperforms non-OA for Twitter, Wikipedia and News in lumbar spine research (Lynch et al, 2022), OA articles obtain a significantly higher Altmetrics score than non-OA articles in knee surgery (Cuero, 2023). OA articles in dermatology received 13.2 mean citations, compared to a matched cohort of non-OA articles with a mean of 7.9, but had no significant difference in Altmetrics Score (Xie et al., 2022). In Library Information Studies (LIS), OA articles were found to have an OAA for both citations and altmetrics (Babu, 2023). OA Climate Change research was found to show an OAA for News (Dehdarirad & Karlsson, 2021). Research into energy technology publications found that OA research was 42% more likely to be cited by Patents than non-OA equivalents (Probst et al., 2023).

## Drivers of the OA Advantages

Potential causes of an OACA have been widely hypothesised, all of which may apply to the OAAA:

1. **Self-selection:** Authors may self-select their best papers to be published OA (Gargouri et al., 2010).

2. **New journals / journal policies:** Key drivers could be new journals, publishing hitherto under published (and potentially poorer quality) research; or well-established journals aggressively promoting OA research.

3. **Acceleration:** OA publication accelerates the rate of scientific production (Woelfle et al., 2011).

4. **Early access:** OA publications receive citations earlier in the publication lifecycle, with the implication that this would disappear over time (Craig et al., 2007).

5. **Improved access:** OA papers receive more citations from researchers without access to subscription journals (Norris et al., 2008; Eysenbach, 2006).

6. **Prestige or network effects:** OA papers that receive attention get more attention than their non-OA equivalents as some form of either prestige or network-based phenomena, similar to the "Matthew Effect" (Ottaviani, 2016; Teixeira da Silva, 2021).

An early study of 27,197 articles published between 2002-7 investigated the 'self-selection' hypothesis and made a number of conclusions: there is no difference in the OACA between self-selected OA and mandated OA research; the highest cited OA articles receive disproportionately more citations; and a possible driver might be user preference to cite the best available research – a form of network effect (Gargouri et al., 2010). A study of 300,000 articles mostly recently published in 2015 showed an 18% OACA, suggesting limited support for the 'early access' and 'improved access' OACA hypotheses, although the former may vary for different forms of OA (Piwowar et al., 2018). The OAAA for five data sources (Twitter, Mendeley, policy, news and blogs) for a small set of UK-based publications published between 2008-13 was observed to persist over time, while policy citations were shown to have an Open Access *disadvantage* in early years; while the news OAAA increased over time (Taylor, 2023). Just as OA publication has been hypothesized to accelerate the rate of scientific production, it has been hypothesized that OA research would increase the creation of intellectual property (IP) as measured by patent citations for drug discovery (Arshad et al., 2016).

## Objectives

As suggested by the above brief review, research into the OACA and OAAA has been limited, whether by locality (Holmberg et al., 2019), journal selection (Wang et al., 2015), sample size (Taylor, 2020) or by discipline (Babu, 2023). This research reports the first complete examination of both the OACA and the OAAA in a very large corpus of all research outputs assigned a DOI and subject area (N = 38.7M), having been published between 2011 and 2021, covering all disciplines and major altmetrics attention sources. The following research questions address the need for extensive and systematic research into the OA phenomenon:

- RQ1 To what extent has the OACA and OAAA phenomena developed between 2011-21, and what are their relationships with the growth of OA publishing?
- RQ2 To what extent do they vary by discipline?
- RQ3 What is the relationship between the OACA and OAAA phenomena, and how do they correlate?

## Methods

The research design was to gather a reasonably complete set of journal articles published between 2011-20 (n = 38,725,313) and to compare both citation counts and altmetrics between the OA and non-OA subsets across six different broad disciplines.

## Data

Journal articles were extracted from the scholarly database Dimensions, which has coverage comparable to Scopus and the Web of Science (Martín-Martín et al, 2011). Dimensions' database is built around metadata harvested from Crossref, Pubmed, Pubmed Central, preprint servers and enriched by publisher supplied data. Articles are classified against Fields of Research categories using machine learning, a process requiring the full text of the article (Hook et al., 2018). In 2019, Dimensions reported that over 100 of the largest scholarly publishers were supplying full-text for the indexing process, and so over two-thirds of the entire corpus are available for analysis; journal articles are further classified by the use of journal-level subject classifications (Herzog et al., 2020).

Only publications matching a minimum of one Field of Research subject classification were retrieved from Dimensions, and these were mapped into six broad disciplines to help identify general trends: Engineering & Technology (ET); Humanities (HU); Life Sciences (LS), Medical & Health Sciences (MHS), Physical & Mathematical Sciences (MHS) and Social Sciences (SS).

Dimensions draws OA data from Unpaywall, a comprehensive OA database (Piwowar et al. 2018); this data is further enriched by inclusion of data from DOAJ, publishers and through indexing additional preprint repositories. Nevertheless, coverage of OA publications will be under-represented, most notably by omitting research without DOIs, and potentially excluding unindexed preprint repositories and publishers; an omission which is likely to offer a bias towards the Global North.

For simplicity, each publication was classified as either OA or non-OA, according to the Dimensions Google Big Data (GBQ) dataset on February 5, 2023. The six disciplines have substantial numbers of journal articles (Table S1), ranging from 36,951 (OA HU 2011) to 845,488 (OA MHS 2021). The OA percentages vary from 21.1% (OA ET 2011) to 63.6% (OA LS 2020). The resultant dataset therefore consists of 132 groups as defined by three factors (year of publication, OA status and discipline).

A dataset from Altmetric was incorporated into a Dimensions dataset, with data from both sources being up to date on February 5, 2023, for Dimensions journal publications published between 2011-21. For each published article, seven indicator values were retrieved: Dimensions' citations and Altmetric's news, blog, Twitter, Wikipedia, patent and policy data. The platform known as Twitter throughout the research period is now known as X; its earlier name is used in this manuscript.

## Analysis

A variety of statistics were calculated for each cohort: for each of the seven indicators, both proportion (i.e., the percentage of the cohort with non-zero activity) and mean values (based on log+1 values) were calculated: these two values were chosen in order to measure the *coverage* of an indicator and the *scale* of that coverage, thus allowing for an analysis that supported the comparison of the proportion getting attention versus the scale of that attention.

For both publication rate and each of the resulting fourteen indicators, the value of the *OA* cohort was divided by its *non-OA* counterpart to achieve a ratio for each indicator quantifying the OAA. The OAA for volume of publications is the Open Access Publishing Advantage (OAPA), the OAA for citations being the Open Access Citation Advantages (OACA proportion and OACA mean), and each of the altmetrics indicators being the Open Access Altmetrics Advantages (OAAA proportion and OAAA mean). Each value shows a positive OAA when >1, and a negative OAA (i.e. an advantage to *non*-OA research) when <1. A value of 1 would indicate that neither OA nor non-OA has the advantage.

To understand whether the annual change in OAPA is significant for each discipline, the breakdown of OA/non-OA for a given year and discipline was compared to the previous year's data points using a Chi-squared population test: this test allows us to calculate the probability that the differences between the two years have occurred by chance.

Analysis of both citation and altmetrics means is rendered complex by data skewness (Hammarfelt 2014; Ottaviani 2016; Thelwall 2017); thus, the appropriate statistical analysis was carried out on the log values of each publications' indicators (all values incremented by 1 to facilitate this computation): the significance of the difference between the OA and non-OA populations being calculated by a Z-test carried out on these log values. Statistical analysis of the population proportions however, is not subject to this skewness, and was analysed by a Chi-squared population test to analyse for variation from the global distribution for that cohort.

To understand the relationship between each cohort's proportion and mean, these values were further analysed by a Pearson test for each indicator; similarly the relationship between each indicator and each cohort's OAA proportion and OAA mean is presented.

## Results

This section presents data covering the growth of journal-based research publications from 2011-21, with a particular focus on the differential growth rates of six different disciplines according to their OA status, trends in citation performance and trends in six different altmetric attention sources (news, blogs, Twitter, Wikipedia, patents and policy documents), with a particular focus on the differential attention rates in six different disciplines according to their OA status. The results, and statistical analysis may be found in the Data Supplement (Taylor, 2024).

## Open Access Publication Trends

Each discipline's publication rates grow substantially over the decade, with an ultimate Open Access Publication Advantage (OAPA) for all but one field. Overall OA rates grew from 32% in 2011 to 55% in 2021. Annual growth for OA proportion approximates to 2% for all disciplines. Growth in OA publishing is more pronounced in disciplines that start from a lower percentage: the proportion of OA papers for Engineering & Technology (ET), Humanities (HU) and Social Sciences (SS) published in 2011 are 21%, 24%, 31% respectively, but have grown to 43%, 51% and 58% by 2021. In contrast, Physical & Mathematical Sciences (PMS), Life Sciences (LS) and Medical & Health Sciences (MHS) grow from 33% to 51%, 45% to 63%, and 36% to 62%. Despite the growth in OA publications, non-OA publication rates in ET, PMS, LS and SS also grow over the same period, albeit at a much lower rate, being between 1.9-1.3 times higher, when comparing 2011 and 2021. Rates of HU and MHS non-OA publishing are static over the decade (Table S1).

In 2013, LS was the first discipline to become majority OA with MHS and SS following in 2016 and 2017 respectively (Figure 1). HU and PMS became majority OA in 2020, leaving only ET majority non-OA in 2021. Analysis of SS and HU disciplines over the period, where year-on-year analysis shows several years with sharp inflection points, strongly suggesting significant non-linear trends in the growth of OA / non-OA publications. In contrast, PMS's transition to OA is considerably slower and showing no significant year-on-year trends between OA and non-OA publications. ET and MHS show 2 and 3 years of non-linear growth respectively.

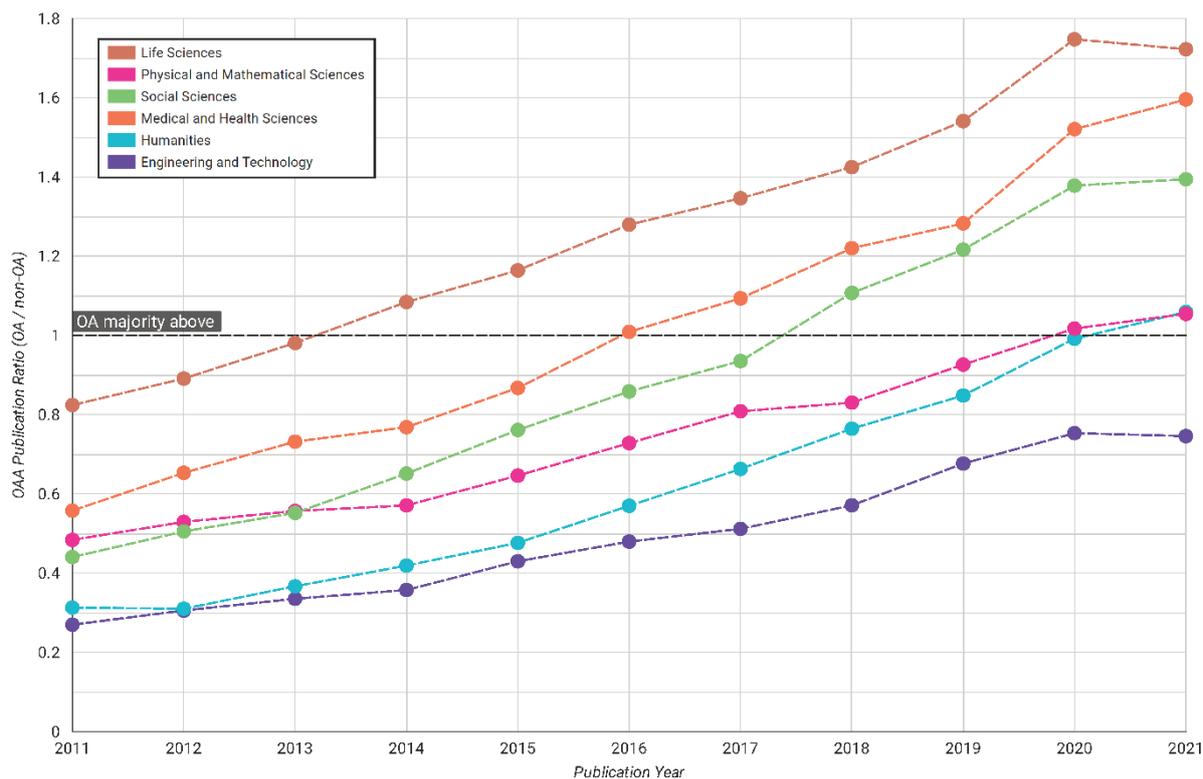

***Figure 1*** *The ratio of OA to non-OA journal articles (OAPA) published per year for journal articles split into six disciplinary groupings. The publication data is from February 5, 2023.*

## Trends in Citations

In general, the OACA proportions (Figure 2) and OACA means (Figure 3) are stronger in Medical & Health Science and Life Science papers than in others. For all disciplines, the OACA means was stronger for older papers than newer papers, while the OACA proportions is largely consistent for MHS and LS for all publication years. For other disciplines, the OACA proportions is largely absent for older papers, and weaker still for newer papers. A strong *disadvantage* for recently published Social Science research is shown. In general, there is a moderately strong correlation between OACA proportion and OACA mean: the correlation is highest for the lowest cited disciplines (Humanities, both OA and non-OA) and weakest for the highest cited disciplines (LS and MHS) (Table S5).

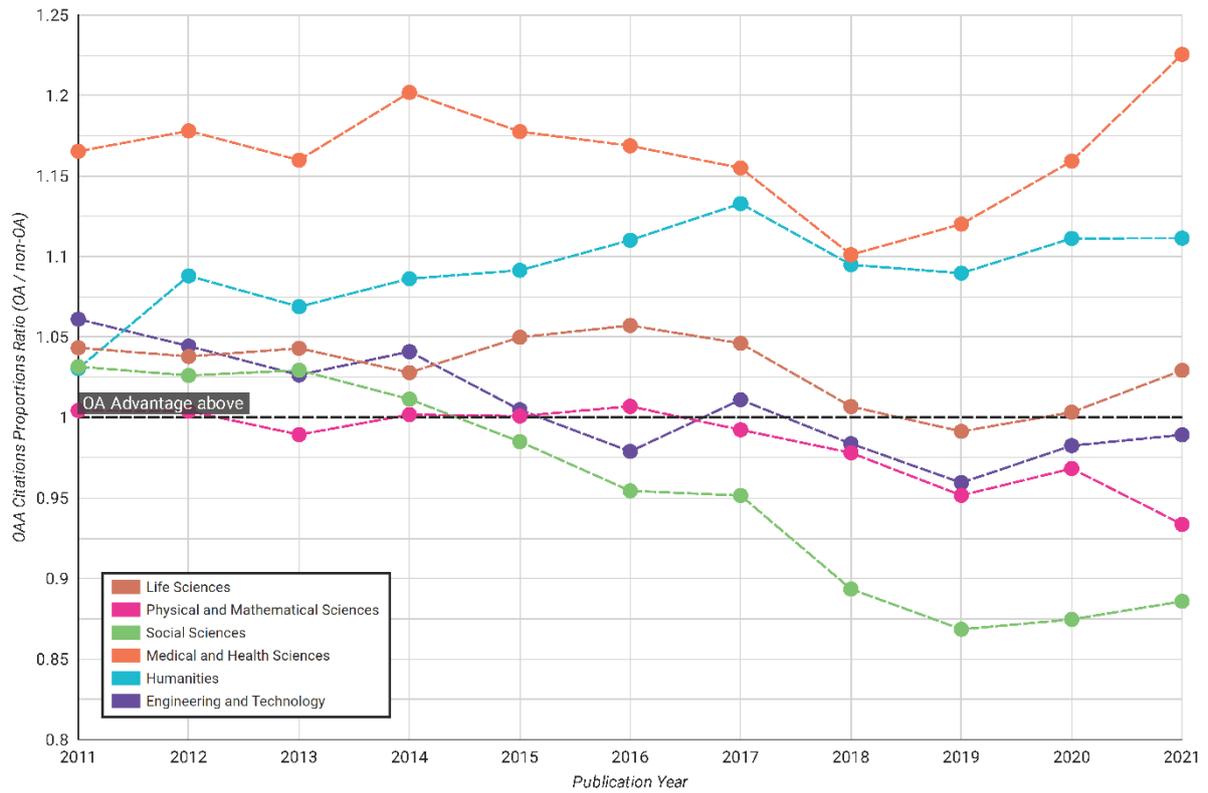

***Figure 2***. *The ratio between the proportion of OA journal articles with at least one citation to the proportion of non-OA journal articles with at least one citation for six broad disciplines in Dimensions. The citation data is from February 5, 2023.*

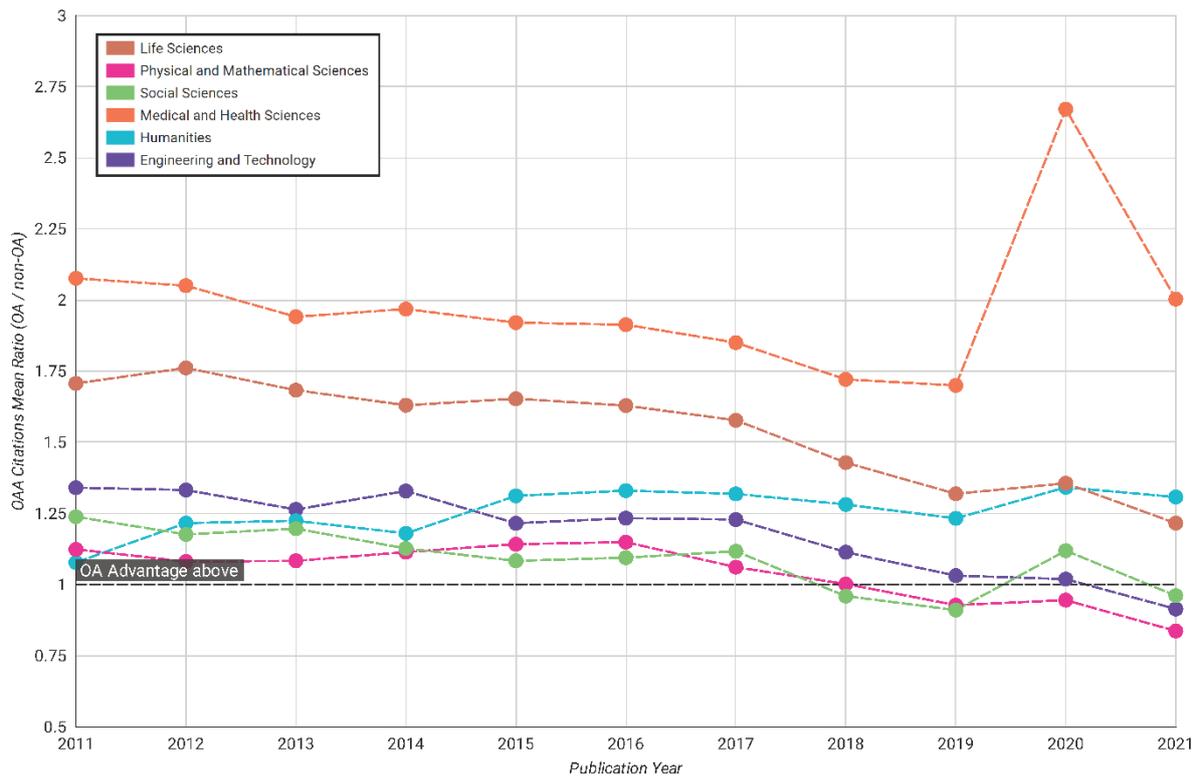

***Figure 3.*** *The ratio between the mean of OA journal article citations to the mean of non-OA journal article citations for six broad disciplines in Dimensions. The citation data is from February, 2023.*

The proportion of papers that receive at least one citation varies over time, by discipline and by OA and non-OA cohorts (Table S3). Physical & Mathematical Science and LS have the largest proportion of papers cited, with over 81% of 2011 papers and over 57% of the newest papers receiving citations for both disciplines. In contrast, only 40% of 2011 HU papers have been cited, and under 15% of 2021 outputs: nevertheless, the OACA proportion for HU research shows the strongest rate of growth over the decade.

As assessed by the Z-test, the citation mean for OA papers is significantly different from the citation mean for non-OA papers for most years when calculated using the logged citation values+1 (Table S4). There are wide variations between disciplines. The most cited sets, MHS OA and non-OA papers in 2011 received a mean of over 33 and 16 citations respectively. In contrast, the least cited discipline (HU) received, on average between 4-5 citations for both 2011 cohorts. The most recently published OA papers for SS, PMS and ET all show a negative OACA for both proportion and mean. As for many other indicators, both means, proportions and OAAAs were higher during 2020-21, the COVID-19 years.

**Trends in News Mentions**

The OAAA proportion (Figure 4) and OAAA means (Figure 5) for news are universally strong: these advantages are stronger for Engineering & Technology, Life Sciences, Medical & Health Sciences and Physical & Mathematical Sciences, while being weaker for Humanities and Social Science papers (Table S4). In general, the correlation between the proportion of articles with news mentions, and the mean of news mentions is very strong, with only HU non-OA research having an *R* below 0.8 (Table S9).

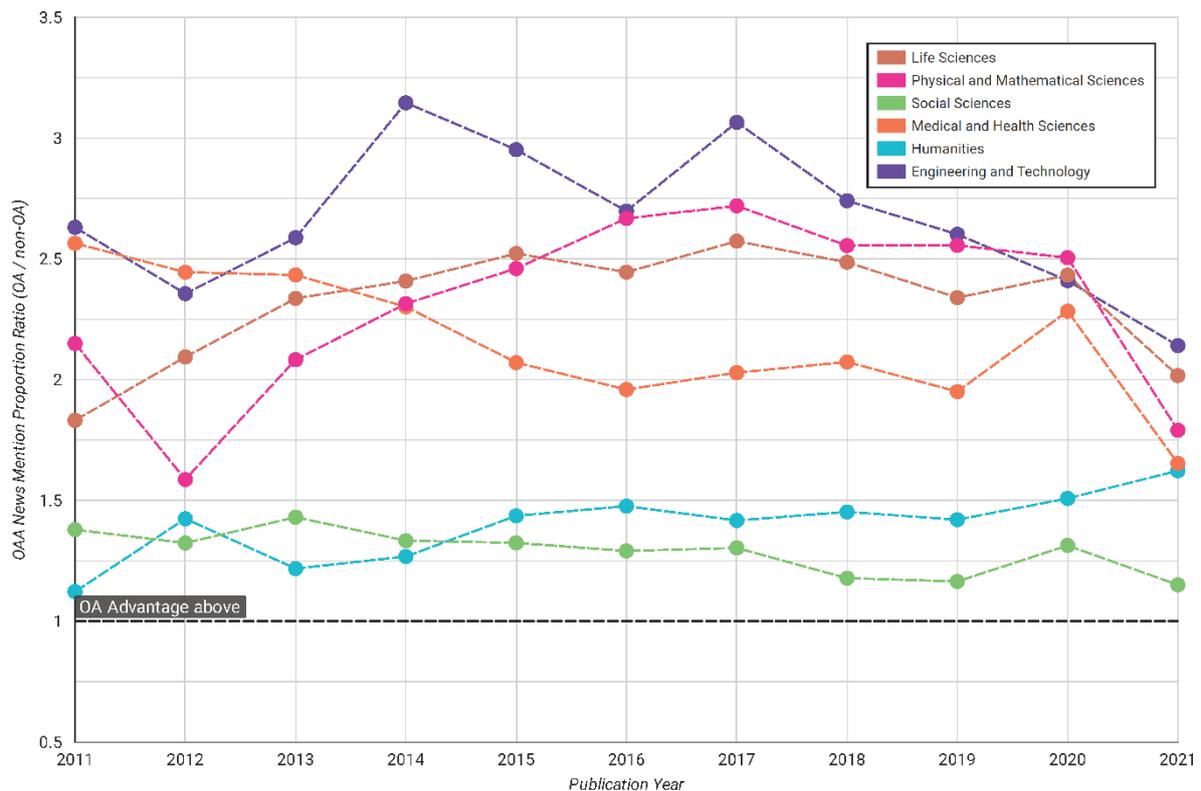

***Figure 4*** *The ratio between the proportion of OA journal articles with at least one news mention to the proportion of non-OA journal articles with at least one news mention for six broad disciplines in Dimensions. The news data is from February 5, 2023.*

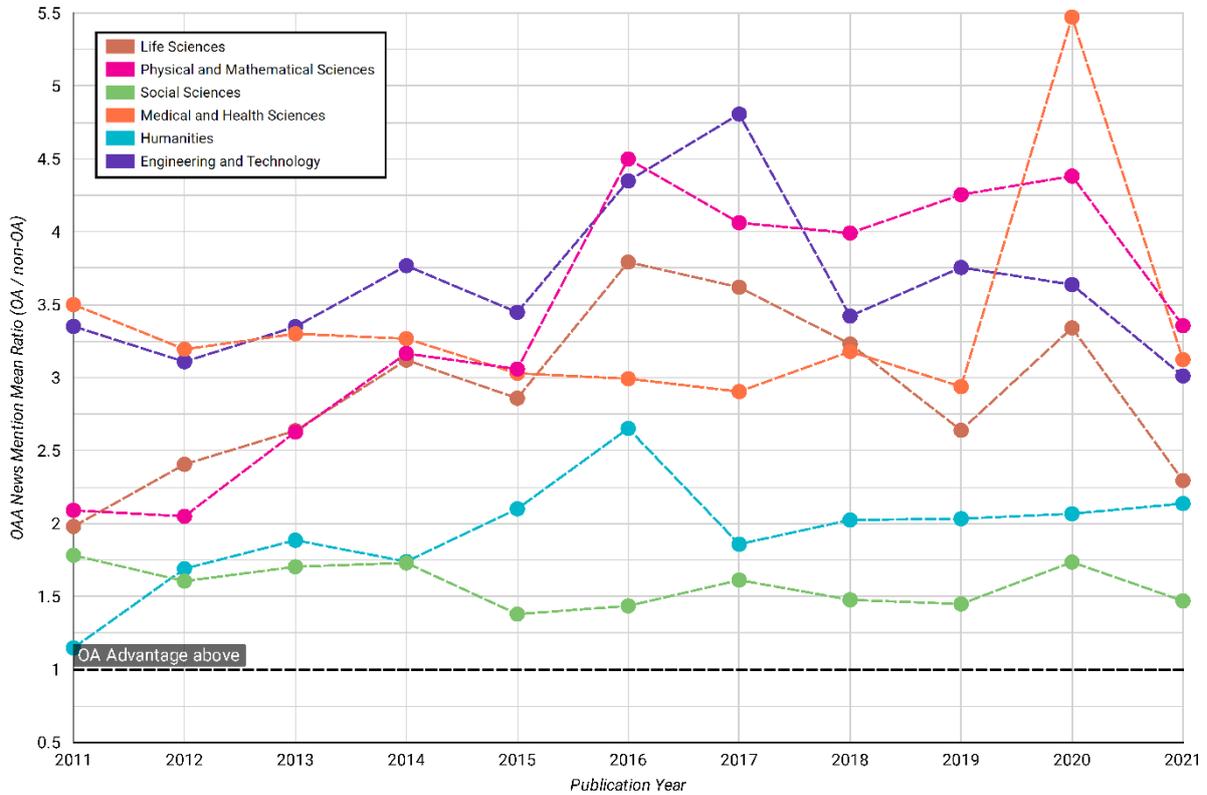

***Figure 5*** *The ratio between the mean of OA journal article news mentions to the mean of non-OA journal article news mentions for six broad disciplines in Dimensions. The news data is from February, 2023.*

There are significant discipline differences for OAAA proportion, with attention being highest for MHS, LS and SS respectively. The proportion of OA MHS papers with news attention range from 3.7% in 2011, to 9% in 2016, and 7.5% in 2021. The fields with the lowest levels of attention are ET and PMS: the former is mostly below 2%, with ~3-4% of OA PMS papers receiving news attention.

ET has the highest OAAA mean for news, being close to 4 for all years. The OAAA mean for PMS rises from close to 2 to over 4, while MHS papers of all years have an OAAA mean between 3-4x, until the COVID-19 years, when it rises to 4.5. LS rises at a similar rate to PMS, although at a lower level. The OAAA mean for HU and SS is generally consistent for papers of all years, with HU having an OAA generally over 2, and SS generally below 2. Despite the low levels of news attention, ET shows the highest OAAA proportion, being about 2.5 through the decade. HU and SS show the lowest OAAA proportions and OAAA means, generally being below 1.5.

## Trends in Blog Mentions

In general, the proportion of papers with mentions in blogs has remained relatively low and static (Table S10), although significant disciplinary differences exist (Table S11). The OAAA proportion (Figure 6) and OAAA mean (Figure 7) for blogs are consistent for publications for all years except for Physical & Mathematical Sciences, with the OAAAs being lowest for Humanities and Social Science publications. The correlation between the proportion of papers with blog attention, and mean blogs varies from moderately to very strong (Table S13).

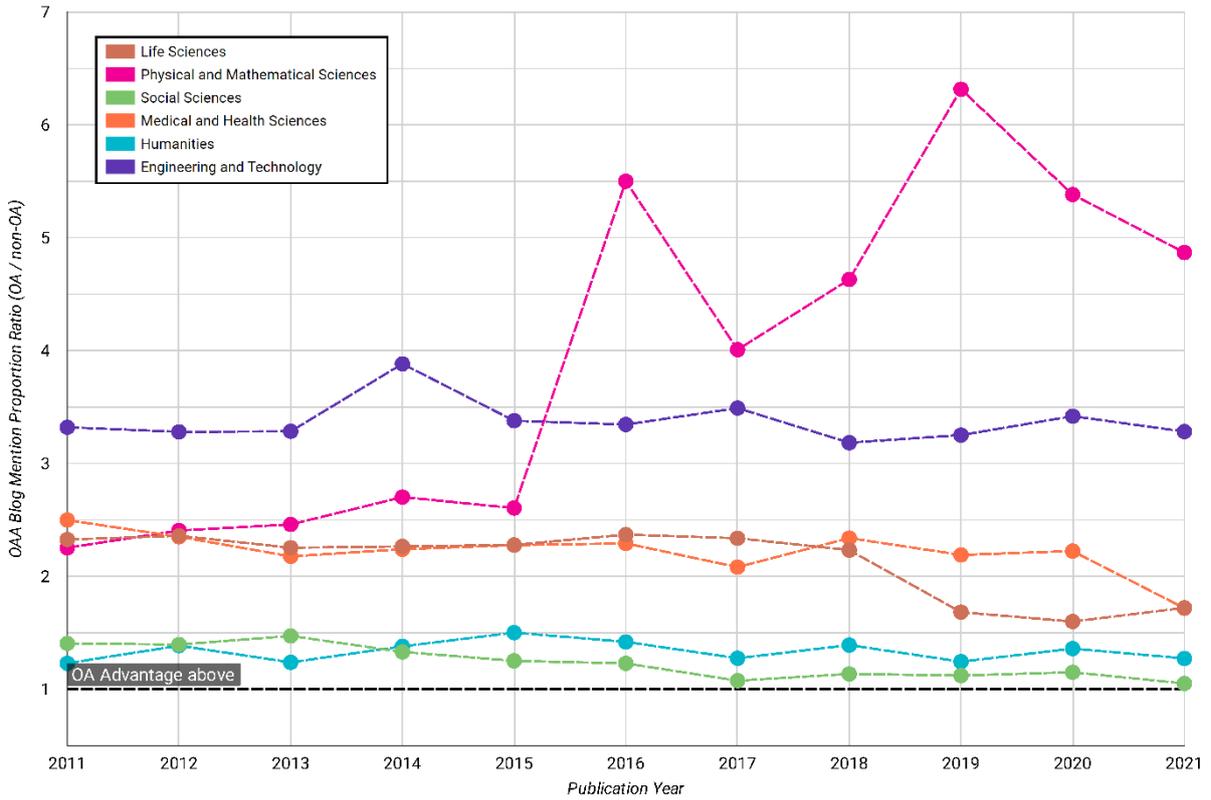

***Figure 6*** *The ratio between the proportion of OA journal articles with at least one blog mention to the proportion of non-OA journal articles with at least one blog mention for six broad disciplines in Dimensions. The blog data is from February 5, 2023.*

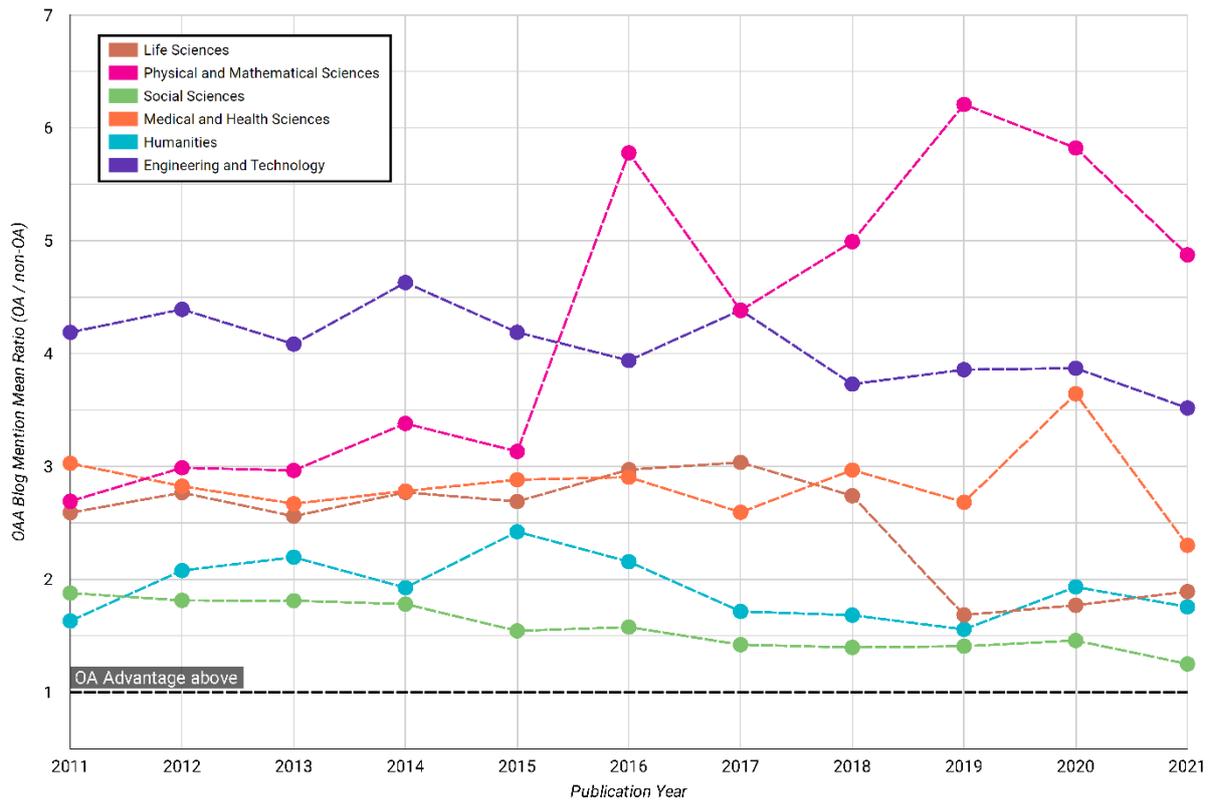

***Figure 7*** *The ratio between the mean of OA journal article blog mentions to the mean of non-OA journal article blog mentions for six broad disciplines in Dimensions. The blog data is from February, 2023.*

The average number of blog mentions to research papers is generally low, with many cohorts being below 0.1, excepting OA LS papers published between 2011 and 2020, OA MHS papers published between 2013-18, and OA SS papers in 2014 (Table S12). The highest proportion of papers receiving attention in blogs were published between 2015-2018 for LS, MHS and PMS, which have peaks over 7%, 4.5% and 6% for each subject's OA publications. Although ET's blog activity is the lowest of the six disciplines (peaking at over 1.5% in 2017 for OA research), it – and LS, MHS and PMS - show a statistically significant OAA.

Physical & Mathematical Sciences show the most striking change: from 2011 to 2015 the OAAAs are consistent, with approximately 2.5x the proportion of OA papers receiving blog mentions over their non-OA equivalents. However in 2016 this proportion jumps to over 5, a level which is more-or-less maintained until 2021, giving PMS the highest OAAA for the proportion of papers receiving attention from blogs: this data was largely driven by the introduction of a semi-automated blogging platform (Arxiver.net), that supported the publication of astronomy preprints.

### Trends in Twitter Mentions

Other than citations, Twitter provides the broadest coverage of research of all sources analysed in this paper. Of papers published in 2011, the lowest proportion of papers with Twitter mentions is 2%, for non-OA papers in Engineering & Technology, the highest is over 16%, for Life Science OA papers (Table S15). The OAA proportions shows considerable disciplinary differences but declines over time for all disciplines (Figure 8). Growth in the volumes of Twitter mentions over the decade is dramatic (Table S16). The OAAA means are high and do not decline for recent papers (Figure 9); almost all observations are statistically significant. The cohort with the least growth is Social Science OA papers, rising from a mean of just over 1 to just over 4, showing a quadrupling of attention. In contrast, the mean values for all LS and MHS papers rise by a factor of approximately 10 between 2011-21. Medical & Health Science OA papers are the most tweeted about, with a mean of almost 16 in 2020, during COVID-19, having been 8 in 2019. OA LS papers also peak in 2020, with a mean of 10. All cohorts show a moderate or strong correlation between the proportion of publications with Twitter attention, and with the mean values (Table S17).

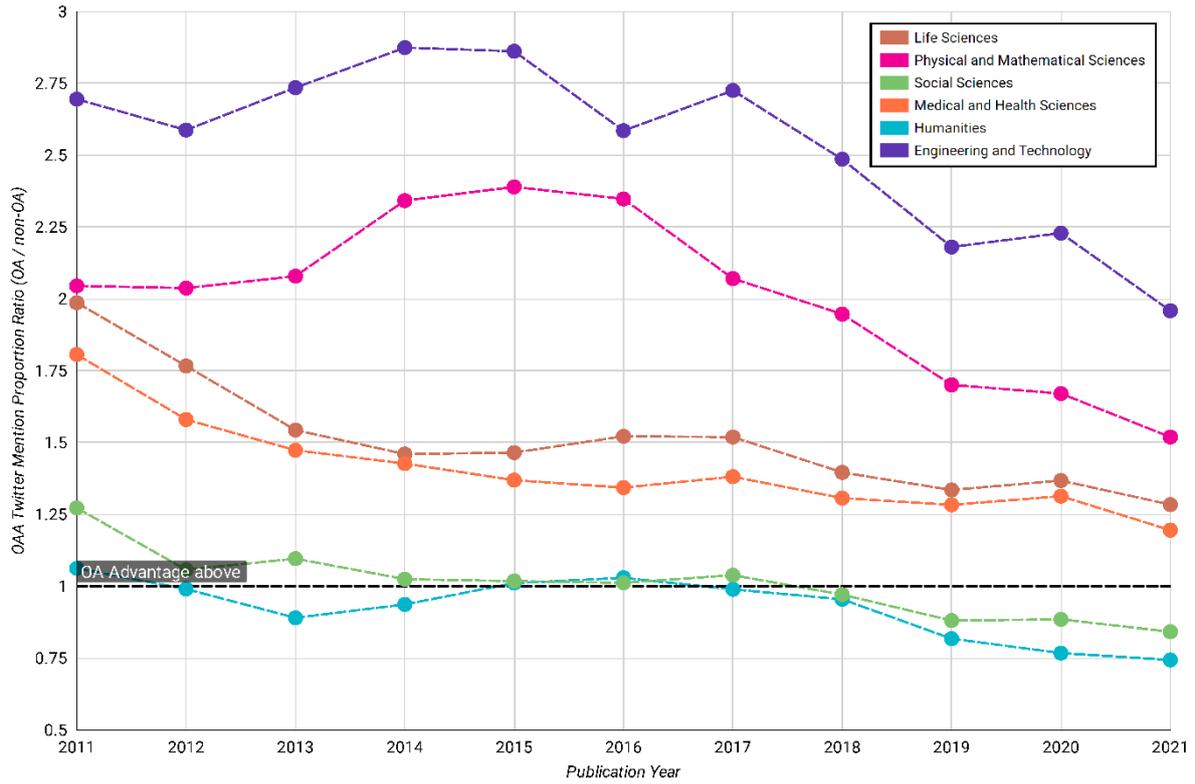

*Figure 8* *The ratio between the proportion of OA journal articles with at least one Twitter mention to the proportion of non-OA journal articles with at least one Twitter mention for six broad disciplines in Dimensions. The Twitter data is from February 5, 2023.*

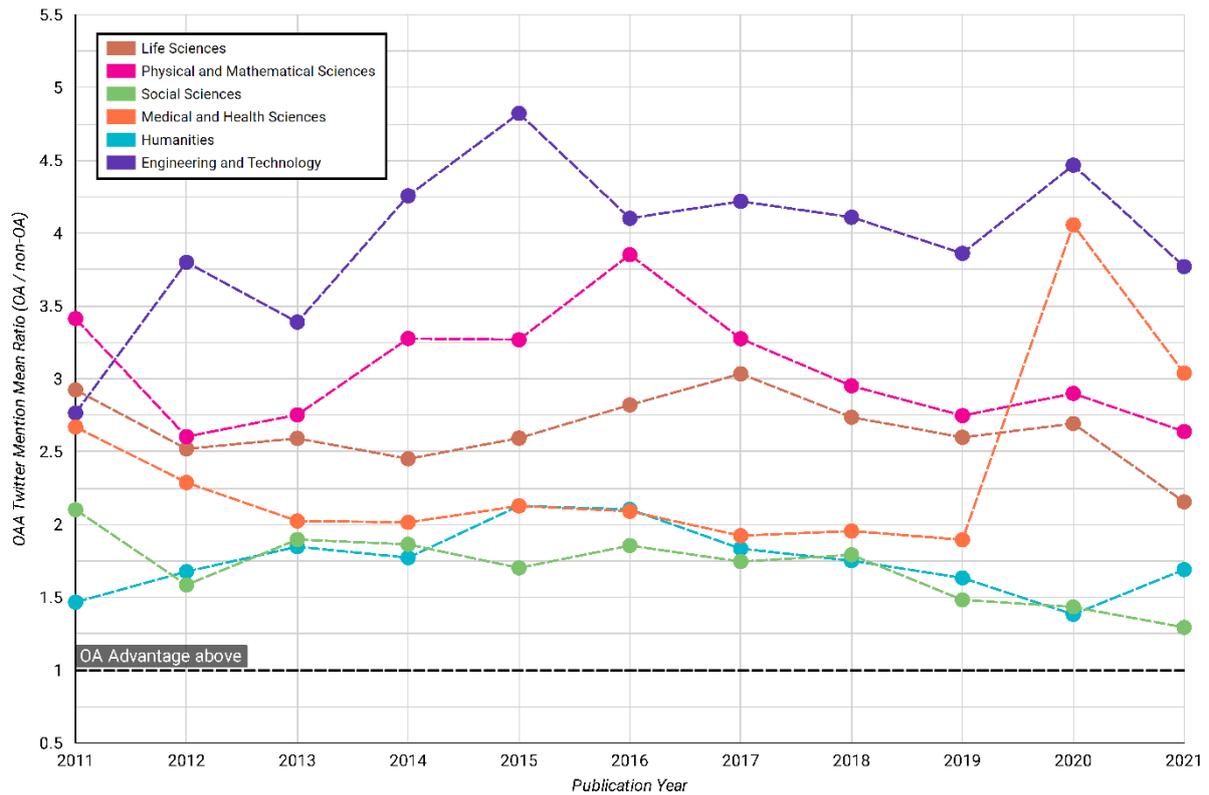

*Figure 9* The ratio between the mean of OA journal article Twitter mentions to the mean of non-OA journal article Twitter mentions for six broad disciplines in Dimensions. The Twitter data is from February, 2023.

For papers published in 2021, the lowest proportion of papers mentioned on Twitter is 11%, for non-OA ET papers, and the highest is almost 50%, for OA LS papers. All proportions grow over the decade, all observations for ET, LS, MHS and PMS are statistically significant. MHS and LS papers consistently have the highest proportion of Twitter attention, for both OA and non-OA papers, with proportions ranging between 11-38% (LS non-OA), 16-50% (LS OA), 11-38% (MHS non-OA) and 20-46% (MHS OA). Social Science research is the third most covered, with a proportion between 11-30% for non-OA papers and 14-30% for OA. ET has the least coverage throughout the decade, peaking at 22% for OA papers in 2020. The OAA Proportion for HU and SS research is largely absent for all years.

## Trends in Wikipedia Citations

The proportion (Table S19) and means (Table S20) of papers with Wikipedia citations is higher for older papers for all cohorts. The correlation between the proportion of papers with Wikipedia citations and the mean value of those citations is very high, with the exception of OA Physical & Mathematical Science papers, which is moderately high (Table S21), and is due to a single outlying paper. The OAAA proportions (Figure 10) and OAAA means (Figure 11) are otherwise consistent and positive for all disciplines except Humanities and Social Sciences, which are neutral. Correlations between proportion and mean Wikipedia citations are very high (Table S21).

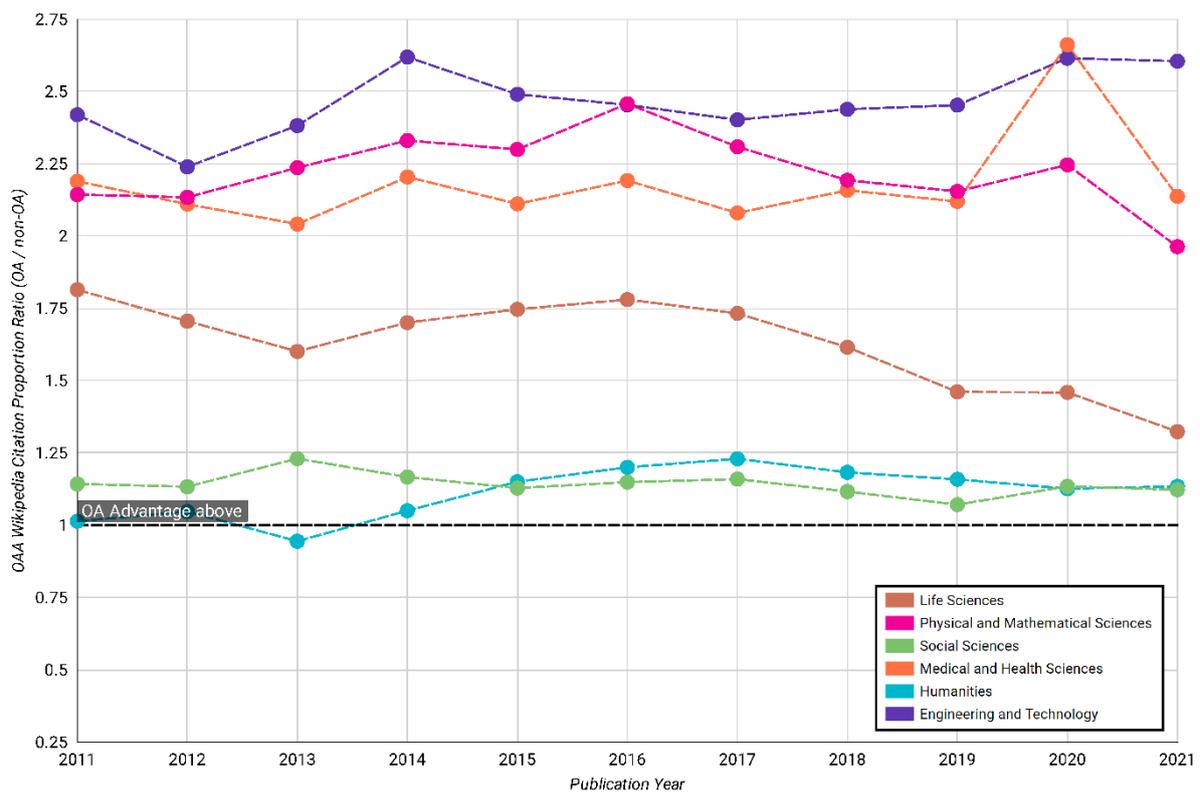

*Figure 10* The ratio between the proportion of OA journal articles with at least one Wikipedia citation mention to the proportion of non-OA journal articles with at least one

*Wikipedia citation mention for six broad disciplines in Dimensions. The Wikipedia data is from February 5, 2023.*

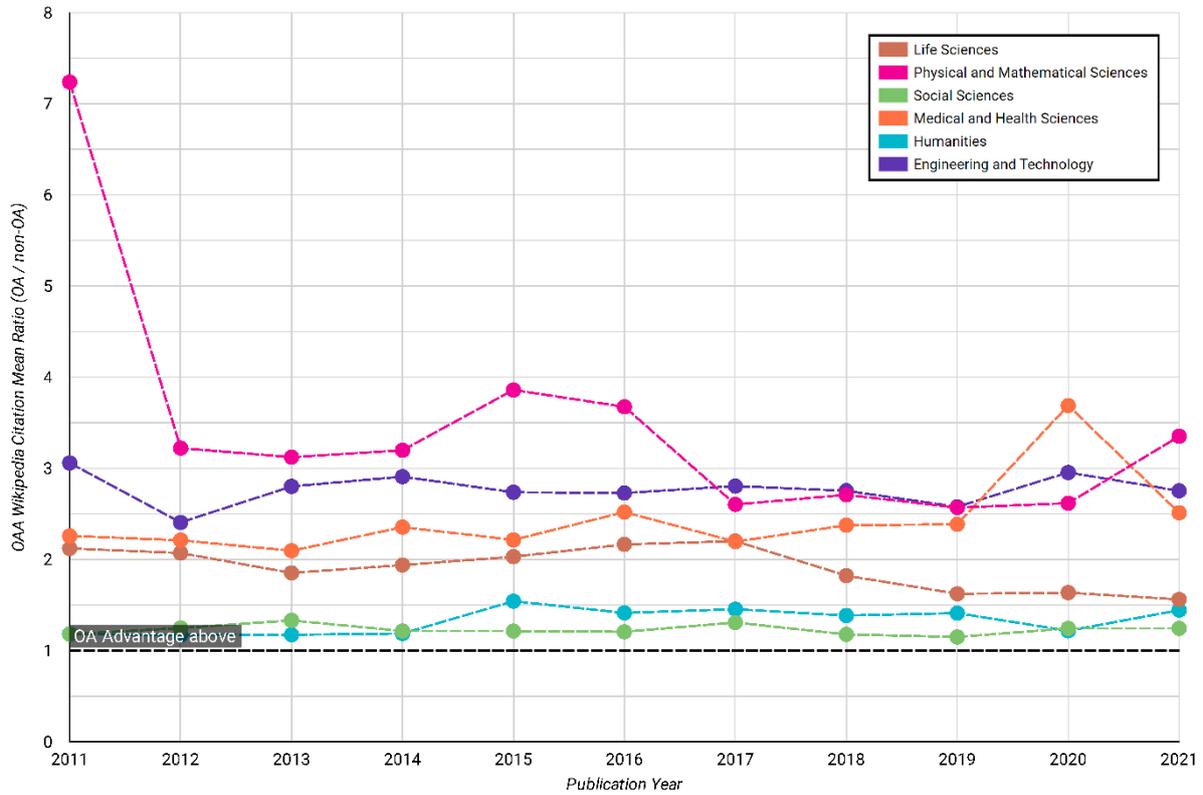

**Figure 11** *The ratio between the mean of OA journal article Wikipedia citation mentions to the mean of non-OA journal article Wikipedia citation mentions for six broad disciplines in Dimensions. The Wikipedia data is from February, 2023.*

Of the papers published in 2011, the discipline with the highest proportion of cited papers is LS, with OA and non-OA papers receiving 7% and almost 4% respectively. MHS OA papers from the same year are cited almost 4% of the time, whereas their non-OA equivalents are cited almost 2% of the time. Fewest ET papers are cited, with only 0.7% of non-OA papers and 1.6% of OA papers getting attention from Wikipedia. In general, the newest papers have received a negligible number of citations from Wikipedia, with the peak values for papers published in 2021 being 0.3 for both OA LS and OA PMS cohorts. The trend by publication year is roughly linear, descending from 2011 to 2021.

ET papers show the highest OAAA proportion, despite their overall low proportion of papers with Wikipedia citations, OA papers are over 2.5x more likely to be cited in Wikipedia than their non-OA equivalents (Figure 10). PMS and MHS papers show a similar advantage, although in the case of PMS, this tails off towards the end of the decade with an advantage of only 2x, and MHS peaks in 2020. The OAAA for LS decreases from 1.7x to below 1.5x over the experiment.

The 2011 value is an outlier caused by a single paper (Masiero et al., 2011), which had received 25,000 Wikipedia citations at the time the data was sampled. This paper reports on the physical properties of several thousand asteroids, and many of these have got a Wikipedia page. Despite N > 180,000, this value contributes ~0.14 to the overall score. The OAAA mean for PMS papers published in 2014 without this value would be approximately 4.

## Trends in Patent Citations

In general, the volume of patent citations is low (Table S24); however the OAAA proportion is significant for Medical & Health Sciences and Life Sciences, the OAAA means are significant for all disciplines and years, except for Humanities and more recently published Social Science papers. Patent citations are heavily skewed towards the older papers, but the OAAAs tend to be consistent for all years. Both proportions and means of papers show considerable disciplinary differences (Tables S23, S24). Correlation between proportion and mean is very high (Table S25).

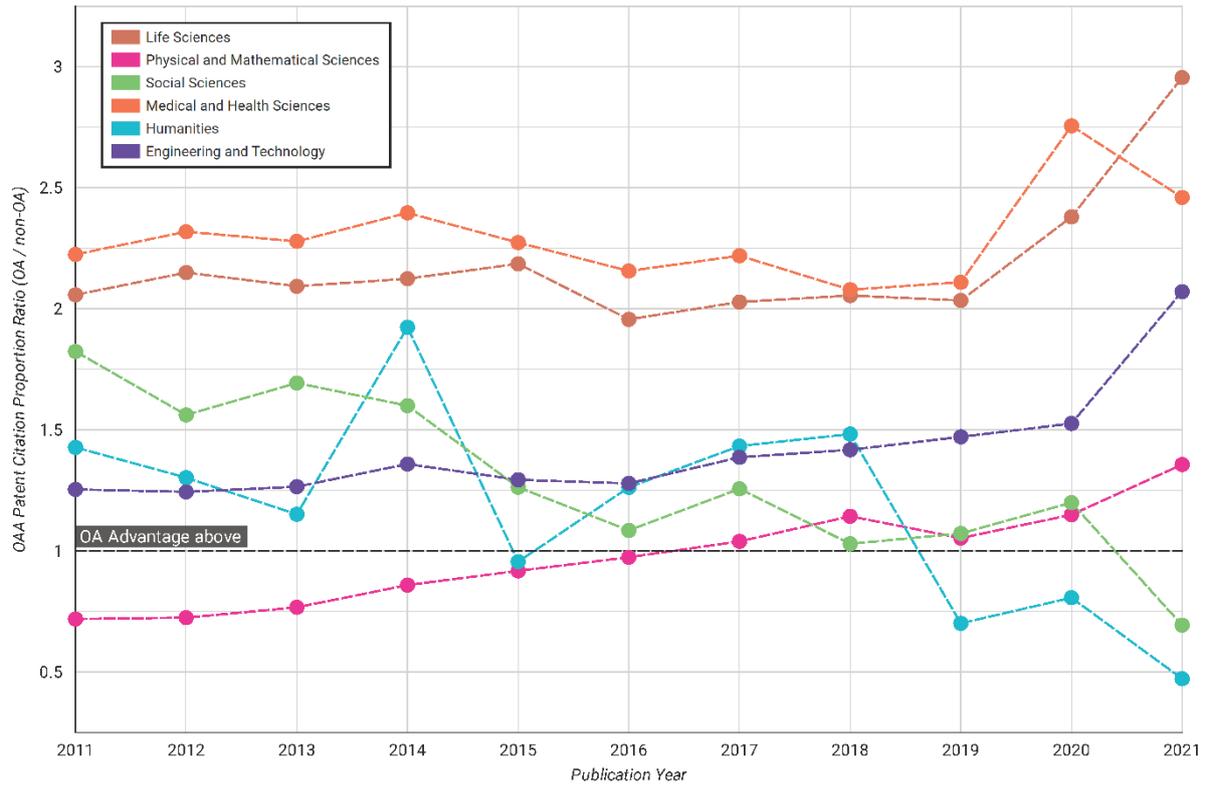

*Figure 12* The ratio between the proportion of OA journal articles with at least one patent citation mention to the proportion of non-OA journal articles with at least one patent citation mention for six broad disciplines in Dimensions. The patent data is from February 5, 2023.

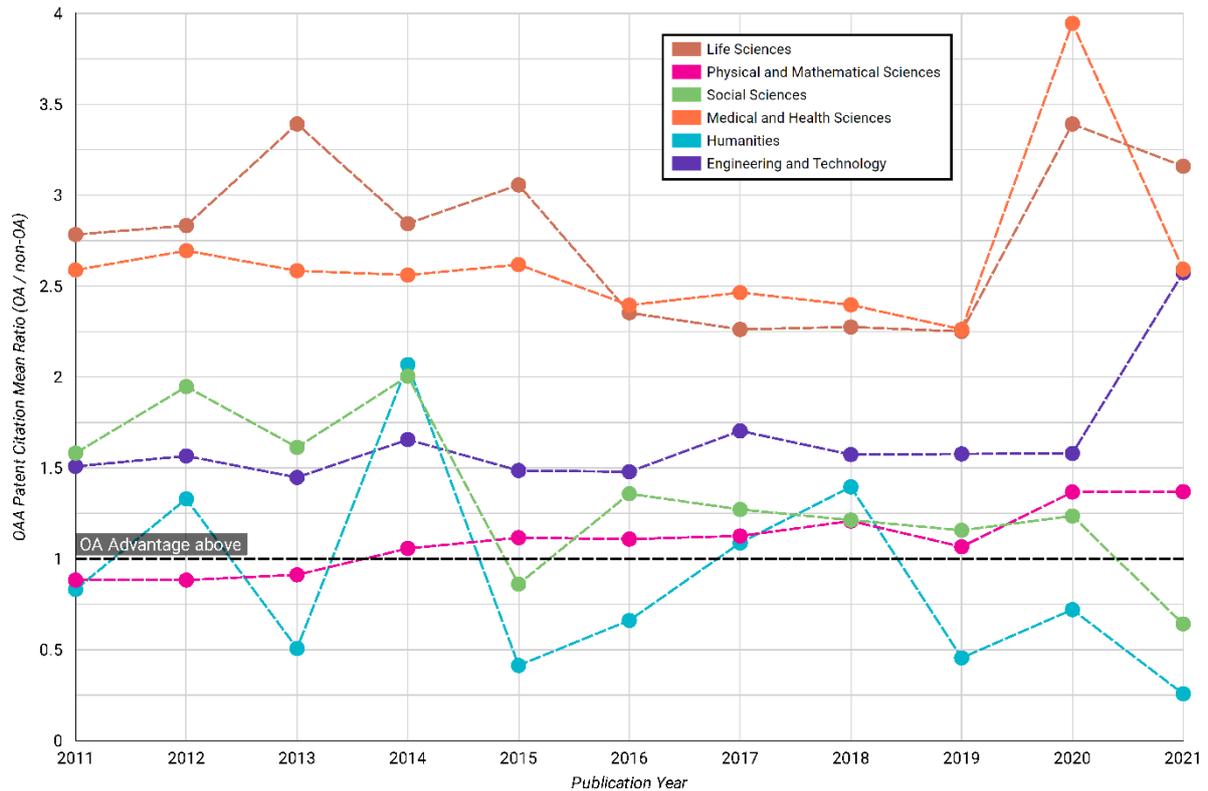

*Figure 13* *The ratio between the mean of OA journal article patent citation mentions to the mean of non-OA journal article patent citation mentions for six broad disciplines in Dimensions. The patent data is from February, 2023*

Although OA LS papers published in 2011 show the highest coverage (over 8% receive patent citations), Engineering & Technology shows consistently high coverage, with over 5 and 6% of 2011 OA and non-OA papers being cited. OA MHS papers from the same year have almost 7% coverage, while their non-OA equivalents have just over 3%. Physical & Mathematical non-OA papers have a higher coverage than the OA cohort, 5.5% versus 4%.

LS and MHS research have a strong OAAA for both proportion and mean, which does not decline for more recently published research. PMS is largely neutral for all years, while ET shows a moderate (but significant) OAAA for both proportion and mean, showing a rise for more recently publications.

LS and MHS papers show significant OAAA means, according to Z-Test analysis, with OA papers consistently receiving 2x the number of patent citations – until 2020, when that advantage increases significantly (Figure 13). PMS shows a significant disadvantage for OA papers from 2011-13, whereas later papers show a small - but significant - OAAA. Apart from 2020, when it increases to over 2.5x, ET papers show a consistent OAAA of approximately 1.5x. Volumes are too low for HU and SS patent citations to form secure conclusions.

### Trends in Policy Citations

The OAAAs for proportions (Figure 14) and means (Figure 15) are generally stronger for newer papers than older ones. Policy citations are amongst the lowest volume of altmetric indicators, and show significant disciplinary differences, with Social Sciences, Life Sciences and Medical & Health Sciences showing the highest values for both proportions (Table S27) and means (Table S28). The correlation between proportions and means is very strong (Table S29).

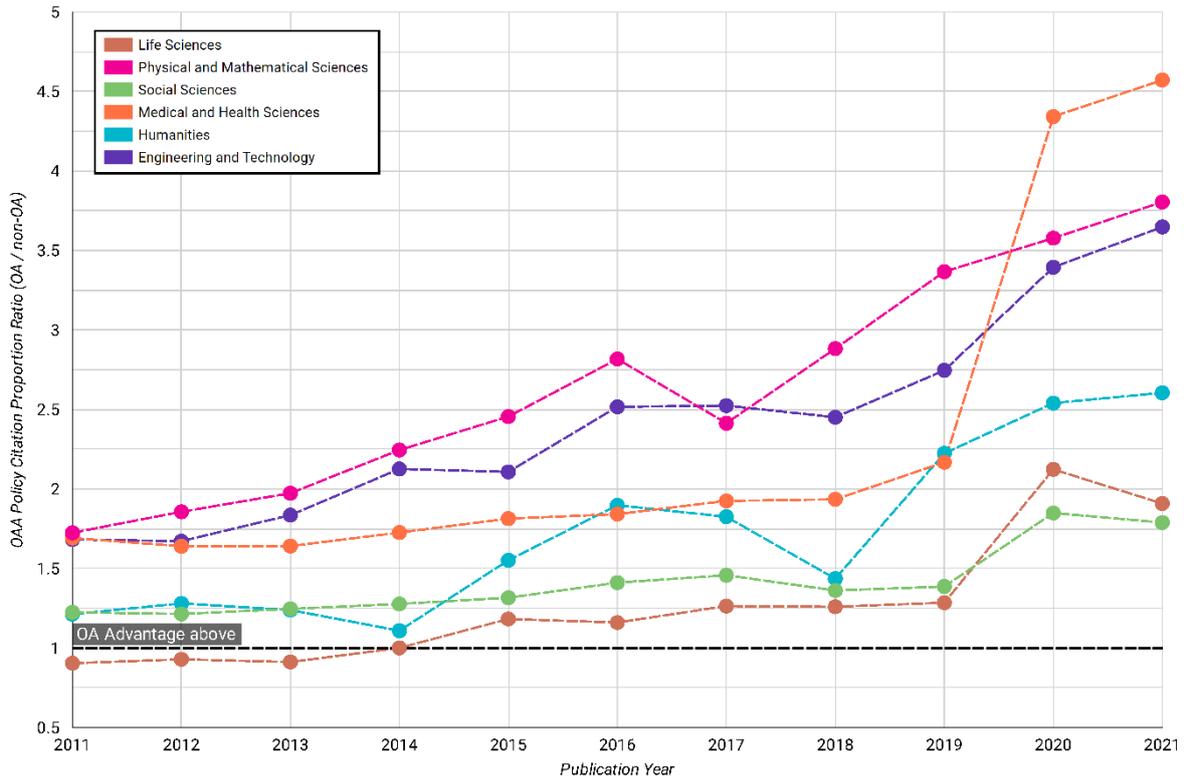

*Figure 14* *The ratio between the proportion of OA journal articles with at least one policy citation mention to the proportion of non-OA journal articles with at least one policy citation mention for six broad disciplines in Dimensions. The policy data is from February 5, 2023.*

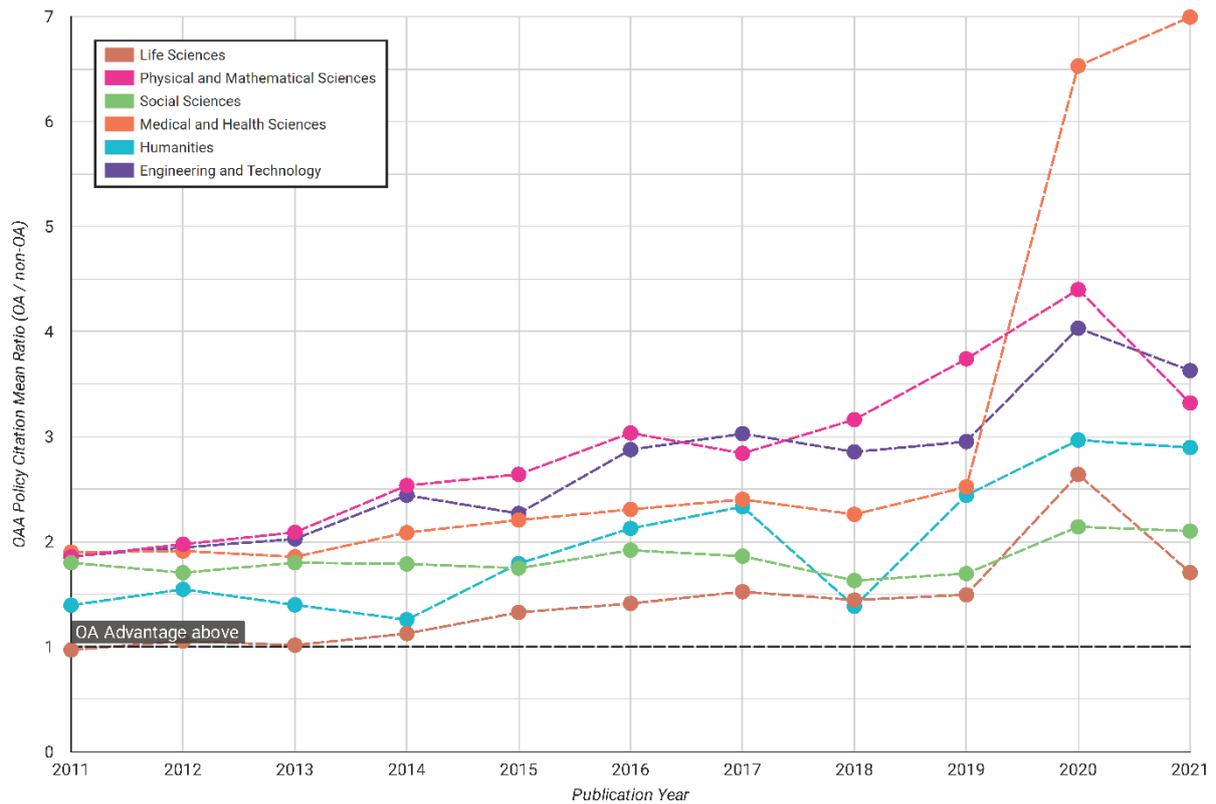

***Figure 15*** *The ratio between the mean of OA journal article policy citation mentions to the mean of non-OA journal article policy citation mentions for six broad disciplines in Dimensions. The policy data is from February, 2023.*

SS papers show consistently show both a high proportion and high means for citations from policy documents, with OA and non-OA papers published in 2011 showing 5% and 4% coverage. Approximately 5% of OA MHS papers published in that year are cited; with between 2-3% of non-OA MHS papers and all LS papers receiving attention from policy documents. PMS, HU and ET have less than 1% coverage for all observations. MHS papers in 2020-21 are over 4x more likely to receive policy citations. Despite the low proportion of papers getting attention from policy citations, PMS papers have the strongest and most consistent OAAA – albeit non-significant.

Despite low volumes of policy citations, significant OAAA are reported for almost all observations, the exceptions being the four oldest LS cohorts, and HU papers published in 2014. In general, despite the lack of significance, the OAAA increases for the more recent papers, with ET and PMS papers showing the higher OAAA values. MHS papers have a mean that is consistently twice that of their non-OA equivalents, until 2020-21, when that value increases to between 6-7x.

## Discussion

This study has a number of weaknesses. Firstly, it is likely that the distribution of research published in journals that haven't been assigned a DOI – and thus not in Dimensions – is not evenly spread by either geography, language or discipline. However, this excluded research is unlikely to be either highly-cited or otherwise impactful, but it is not known whether it has different OACA or OAAA relationships. Also excluded are papers which cannot be assigned a subject area, which will likely be biassed towards smaller publishers. Secondly, this research amalgamates all forms of OA publishing, and thus cannot report on the relative success of the different models. In particular, differences over time may be due to the emergence of large OA journals from traditional publishers, such as Springer, and the rise of large OA publishers such as MDPI and Frontiers, as well as to varying proportions of predatory publisher articles. Thirdly, this study is not a longitudinal study: research published in 2011 and reported here has had 13 years in which to accrue attention, whereas research published in 2021 has had only three. An additional weakness is that research is known to switch between non-OA and OA. Each paper in this research has only one status, that which was reported by Unpaywall at the time of sampling.

## The development of the OAAs between 2011-21

The increase in OA publication was faster than non-OA, resulting in all disciplines, except Engineering & Technology, becoming majority OA by 2021. These results are consistent with previous investigations: the 20% reported in 2009 (Björk et al., 2010) is in line with 32% presented here for 2011; 41% in this paper for 2015 versus 45% (Piwowar et al., 2018) and 55% OA for 2021 is in line with 54% previously reported for 2022 (Basson et al., 2022).

For RQ1, the OACA proportion and OACA means are broadly steady across the decade, a slight decline is reported. Although these results correspond with earlier research (Piwowar et al., 2018), the differences between OACA proportion and OACA mean, and between disciplines have not hitherto been reported.

OAAAs do not follow the same patterns as OACAs. The OAAAs for news and blogs are largely consistent for each publication year, with the implication that attention paid to OA research by these sources is independent of the volume of OA publication. These broader observations for news and blogs are consistent with earlier, narrower research (Lynch et al, 2022; Taylor, 2023; Cuero, 2023), although covering a wider timescale, encompassing a much larger set of publications and analysing the differences between proportion and mean. Twitter OAAAs partially follow these trends: the OAAA proportions for more recent papers are slightly lower, following a similar pattern to the OACA proportion. However the OAAA mean for Twitter is significant and – with the exception of the COVID-19 years – largely stable, reinforcing earlier, narrower observations (Adie, 2014; Taylor, 2023). While the OAAA proportion declines, the OAAA mean is largely unaffected by the transition to OA: this research adds and extends earlier findings relating to the growth in the use of the Twitter platform (Holmberg & Thelwall, 2014; Mohammadi et al., 2018). Wikipedia and patent citations exhibit consistent OAAAs for both means and proportions for both older and newer papers, despite being amongst the slower metrics to accrue. These results are compatible with earlier, more limited findings for Wikipedia citations (Teplitskiy et al., 2017) and patent citations (Nishioka & Farber, 2020; Probst et al., 2023). The OAAAs for policy citations, however, are larger for more recent papers than their older equivalents, despite the relative slowness of this attention source: this finding expands and elaborates early longitudinal observations (Taylor, 2023).

## Disciplinary Differences

The results on OA publication trends presented here are in-line with smaller studies and confirm the uneven adoption of OA publishing across disciplines (Frantsvåg & Strømme, 2019; Ko et al., 2009; Laakso & Björk, 2022). HU and SS both show low proportions of OA in 2011; however both disciplines have years where the proportion of OA significantly grows against the background trend of overall growth, possibly as a consequence of funding policy or the launching of new, OA-first journals. Consequently, whereas Humanities and ET show a similar OA proportion in 2011, the position changed, so that by 2020, HU became majority OA, while ET remained majority *non*-OA. Neither Life Science (majority OA in 2013-14), Medical & Health Science (2016) nor Physical & Medical Science (2020-21) have years with similar jumps towards OA majority, showing smooth growth over the decade.

For RQ2, this research reports a number of novel findings on the performance of disciplines, and where they are seen to differ. Disciplinary differences have been widely reported, for both citations (Moed et al., 2012) and altmetrics (Fang et al., 2020; Holmberg et al, 2020), and the observations made here expand knowledge of this phenomena and elaborate on them, to provide unique insights into the scope and scale of the various OAAs over time, and over discipline. Although the general trend for the OACA proportion to drop between 2011-21, the OACA proportion for LS and MHS grows against trend, whereas SS research drops markedly by the same metric, showing a *disadvantage* for OA papers for more recently published research. In contrast, the OACA mean generally drops smoothly for papers published across the decade. The one exception is citation behaviour for MHS during the COVID-19 years, which corresponds to work reporting the growth of COVID-19 related OA research (Park et al., 2021). During 2020-21 OA papers in 2020 receive over 2.5x the citations compared to their non-OA equivalents, presumably due to early COVID papers being highly cited and allowed to be published OA by their journals.

The OAAAs reported for news and blogs show a clear difference between HU, SS and the other disciplines, with much stronger OAAAs being reported for LS, MHS, PMS and ET. The OAAA mean for news is universal, persistent and generally steady over time, varying from 2 (HU and SS) to 3-4 times (ET, LS, MHS, PMS). Similar phenomena and disciplinary differences are seen for blogging. These broader observations are in-line with earlier, niche research (Holmberg et al, 2020; Dehdarirad & Karlsson, 2021; Lynch et al, 2022; Taylor, 2023; Cuero, 2023). These differences are somewhat echoed by Twitter OAAs. In each case, the disciplines with the highest OAAAs are ET and PMS, fields that show lower rates of OA publication. Wikipedia citations show similar OAAAs, despite the relatively high rates of HU and SS research citation. Patent citations have the highest OAAAs for MHS and LS, despite relatively high rates of citations for PMS and ET, the OAAAs for patent citations are comparatively low, and broadly in line with SS and HU, disciplines that have almost no patent citation data. Despite significant differences in discipline performance for policy citations, the OAAAs are remarkably similar, all of which grow over the decade – in contrast to all other indicators.

## Comparison with OACA and OAAA hypotheses

For RQ3, we can endeavour to compare the different metrics, and offer potential interpretations against the various hypotheses of the OAAs. This research covers a period starting in 2011, when 32% of research was OA across a period when OA increased to 54% (Table S1). For the same publications, we see the OACA mean dropping slightly for all fields (except for the COVID-19 years), with the most strongly OA disciplines – MHS and LS – showing the strongest and most consistent OACA. The OACA proportion however, rises for HU, remains stable and strong for MHS, is neutral for PMS, LS and ET, and decreases for SS.

If it were the case that researchers were self-selecting their best papers to be OA (Gargouri et al., 2010), the growth of OA over the decade would likely result in a drop in both OACAs as the OA share increased, and this would be most marked for the disciplines that had made the biggest change. The presented data does not provide evidence for this hypothesis: the subjects with the largest growth in OA publishing – LS and MHS – show the highest OACAs, even allowing for the COVID years. ET - which shows the lowest level of OA publishing from 2012-21 - shows no citation proportion advantage, and one of the smallest citation average citations, while its nearest neighbour, HU, has the strongest growth in OACAs.

However, the observations for LS and MHS are consistent with the hypothesis of accelerated scientific production (Woelfle et al., 2011): for these disciplines we see a growth in OA publishing, and with that increase in volume, we see increases in the likelihood of any citation, and increases in the mean citations for those OA publications; a similar trend can be discerned for HU research. The slow growth of OA in ET and PMS is matched by a lack of OACAs. In contrast, data for the SS suggests a different interpretation: here, we see one of the highest rates of migration from non-OA to OA, coupled with the lowest OACA for proportion, and a negligible effect for citation means. For this discipline, the transition to OA has been the third fastest, but has resulted in fewer of those OA papers receiving any citations whatsoever, permitting the possible interpretation that OA publishing in the SS cohort has resulted in a disproportionate growth of poor quality research (Beall, 2015).

In terms of news, blog and Twitter activity, we see consistent OAAAs for most measures and disciplines, with SS and HU publications having much smaller OAAA values than the other disciplines. These OAAA values do not appear to be related to either OAPA or OACA, suggesting different mechanisms behind these phenomena: while there is evidence to support the hypothesis that 'the rich get richer' (Ottaviani, 2016) for HU and SS on Twitter (the OAAA proportion is absent, the OAAA means is present), the general trend is that OA publications are always more likely to benefit from increased attention once visible on these platforms, and – once visible – are more likely to get attention from the same source, a form of network effect (Teixeira da Silva, 2021). A possible additional explanation is the increasing use of the Internet to mediate communication is seeing an expansion of the capacity to transmit research findings, one that is further enabled by OA research.

For patents, Wikipedia and policy citations the data is skewed towards older papers, i.e. years without an OAPA, hence the data is more challenging to interpret. The slow development of the OAAA for policy citations suggests a different mechanism for the translation of policy impact: an interpretation might be that policy authors are accustomed to using a certain set of high profile journals that come from a non-OA tradition, and that these are amongst the slowest to adopt OA. Certainly, the widespread phenomena of publishing COVID-19 research as OA can be seen in both proportion and mean metrics (Park et al., 2021).

The findings for patent citations, and their correlation with OA trends are potential evidence for the hypothesized increase in IP creation (Arshad et al., 2016); alternatively, the consistency of the patent OAAAs for MHS and LS suggests a different mechanism, which could be explained by improved visibility: it is unlikely to be driven by quality, as there is no clear evidence of either OAAA being influenced by the OAPA of the disciplines. This linearity is shared with the OAAAs for Wikipedia: here it has been suggested that the use of an OA 'bot' – that helps complete OA references, and thus providing a visibility advantage – might be partially responsible for an OAA (Holmberg & Vainio, 2018). The latter hypothesis can be seen as a form of 'free access' – the Wikipedia editor can cite OA materials more easily, and so does. Similarly, the explanation might hold true for patent authors, if it could be shown that they don't have access to paid-for journals. The observation of increasing OAAAs for policy citations requires another explanation, this data may show the increasing familiarity of policy authors with OA research, and indicate their slow transition away from a tradition of citing long-standing journals of quality.

## Conclusions

This research sheds light on the development and performance of OA research throughout the key years of 2011-21, during which time OA papers grew to dominate non-OA research by volume. This change has not come about by chance, having required multiple governmental and non-governmental initiatives; neither is it cheap, having an estimated annual cost of $2B. Nevertheless, OA publications are often associated with higher impacts than non-OA research, both in terms of academic impact, as measured through citations; presence in public visibility and discourse in news, blogs and on Twitter; and largely demonstrate increased socio-economic impact, as measured by Wikipedia, patent and policy citations. However, the adoption of OA is far from consistent across disciplines, and – possibly as a consequence – the apparent advantages that OA research often demonstrates may not be shared by these other fields. Furthermore, the advantages of OA are not evenly distributed: while there is

evidence that some fields (Medical & Health Science, Life Sciences, Humanities) are being strengthened by OA adoption, there is the possibility that others (Social Sciences) are being weakened. Additionally, it is notable that while some fields appear to have their visibility and socio-economic impact boosted by their OA status, others (Humanities, Social Sciences) are not similarly benefited. OA status is therefore not the only confounding factor contributing to the impact of research: OAAs are complex, dynamic and multi-factorial, and require considerable analysis to understand and predict.

## Data availability and authorship statement

The data analysed in this paper is available in the Data Supplement. Michael Taylor designed the experiment, analysed the data and wrote the manuscript. The author is an employee of Digital Science, and received funding for the OA APC fee. The data supplement is published and cited, and can be downloaded from Figshare (https://doi.org/10.6084/m9.figshare.25036754.v2)